\documentclass[a4paper,10pt]{article}
\usepackage{float} 
\usepackage[round]{natbib}
\usepackage{enumitem}
\title{A Revised Publication Model for ECML PKDD}
\author{Hendrik Blockeel \and Kristian Kersting \and Siegfried Nijssen \and Filip Zelezny}

\begin{document}

\maketitle

\newcommand\hendrik[1]{{\bf *** hendrik: #1}}{}

\begin{abstract}
ECML PKDD is the main European conference on machine learning and data mining. Since its foundation it implemented the publication model common in computer science: there was one conference deadline; conference submissions were reviewed by a program committee; papers were accepted with a low acceptance rate. Proceedings were published in several Springer Lecture Notes in Artificial (LNAI) volumes, while selected papers were invited to special issues of the Machine Learning and Data Mining and Knowledge Discovery journals. In recent years, this model has however come under stress. Problems include: reviews are of highly variable quality; the purpose of bringing the community together is lost; reviewing workloads are high; the information content of conferences and journals decreases; there is confusion among scientists in interdisciplinary contexts. In this paper, we present a new publication model, which will be adopted for the ECML PKDD 2013 conference, and aims to solve some of the problems of the traditional model. The key feature of this model is the creation of a journal track, which is open to submissions all year long and allows for revision cycles.
\end{abstract}

\section{Introduction}

We propose a new conference/journal publication model, the main feature of which is the creation of a journal track that allows all-year submissions to the conference. We would like to adopt this model at ECML PKDD 2013, but hope it will be useful beyond this conference.  We start this text with arguments why a new model is needed.  Next, we summarize the goals we wish to achieve, and the basic ideas underlying our model.  We end with a detailed description of how we intend to implement it.

\section{Motivation}

Computer Science is atypical as a scientific field, in that it focuses on publishing at conferences, rather than journals.  
There is a growing discontent, internationally and in various subfields of computer science, with this tradition.  Many argue that the conference-oriented publication system has reached its limits, and is breaking down.  See the Appendix~\ref{arguments} and the bibliography for detailed arguments. In particular the ideas of Halpern and Parkes (2011) are very similar to ours.  Briefly, in its current form, the system leads to reviews of highly variable quality; proliferation of conferences (such that the main purpose of conferences, bringing the community together, is lost); high reviewing workloads; slower reviewing for journals; decrease of information content of conferences and journals alike; lengthy journal articles; confusion among scientists in interdisciplinary contexts; unfair evaluations of academics.

The model adopted in most other domains of research, where conferences serve as community meetings and journals as the main publication channel, suffers less from the above shortcomings.  Full papers are reviewed more thoroughly, with multiple reviewing rounds when necessary; there is less confusion about the expected standards; there is no artificial limit on the number of submissions that can be accepted; the total reviewing workload is lower (fewer different versions of papers are reviewed); journal articles are more concise (they need not ``significantly extend'' previous full-length versions).  The main problem with journals, in computer science, seems to be the processing time: journal reviewing tends to take much longer than conference reviewing.

There have been several initiatives to address these issues.  ECMLPKDD has experimented with selecting papers for exclusive publication in journals; this resolves some of the mentioned issues, but not all. SIGGRAPH and ICLP have taken similar initiatives to directly publish conference papers in journals. IJCAI'13 will have a track in which Journal of Artificial Intelligence Research articles will be presented that were not published in any form at a conference. Several conferences (AAAI, IJCAI, ECMLPKDD'12) are introducing NECTAR tracks in which important papers published in related communities can be presented. VLDB has moved to a journal-reviewing-like model.  Changes towards different publication models have been considered for ICML and NIPS, but have not been implemented up till the 2012 editions. ACML'12 has introduced a system with two yearly submission deadlines for the conference track, allowing for resubmissions. The ACL conference on computational linguistics has moved to the VLDB system from 2012.

%The weaknesses of the conference publication system were recognized by earlier organizers of ECMLPKDD. This led to an experiment in which the ``best'' conference papers were immediately included into a journal. This system addressed in particular the ``conference versus journal'' dilemma, and aimed to increase the reputation of ECMLPKDD and the ML and DMKD journals outside computer science. At the same time, this experiment reveiled multiple difficulties: (a) a general discontent with the quality of journal papers published in the special issue, as the conference reviewing process turned out to be not reliable enough; (b) the selection of only a small set of papers that are ``of journal quality'' suggested that the rest is not. 

\section{Goals}

The model that we propose for ECML PKDD 2013, aims at achieving four main goals.

\begin{enumerate}
\item {\bf Further increase the quality of full paper presentations at the conference}, by improving the quality of reviewing and providing more visibility to top contributions at the conference. 

\item {\bf Further increase the quality of journals}, by evaluating a new model for journal reviewing that combines quality with speed.  

\item {\bf Make the conference more inclusive}, by allowing for a broader set of contributions to be presented at the conference, albeit not as journal publications.

\item {\bf Reduce the ``conference versus journal'' dilemma} that authors face. 
\end{enumerate}
Even though our goals are ambitious, we would like the implement the new system by the smallest possible changes to the current ECMLPKDD system.  These changes are
\begin{enumerate}
\item {\bf safe}: the current procedures for conference and journal reviewing remain in place, we only install an alternative track; 
\item {\bf small}: the alternative track is quite similar in spirit to previous ideas of selecting papers for inclusion into a journal directly; \item {\bf advantageous}, both to the ECML/PKDD conference as to the journals taking part in the initiative (MLj, DMKD).
\end{enumerate}
The model is similar to the VLDB publication model, which is increasingly known within the computer science community.

\section{Concepts of the Proposed System}
The starting point of our proposal is that journal reviewing systems are the standard for high quality
reviewing. 
%We would like to impose a reviewing system of journal quality on (some) conference submissions, hence leaving no doubt that these publications are of journal quality.
The journal system has the following key 
distinguishing features:
\begin{itemize}
 \item articles can go through {\bf iterations of revisions} before being accepted;  
 \item there is {\bf no submission deadline}: articles can be submitted, evaluated, and accepted all year long;
 \item reviewing for journals is typically done by {\bf senior researchers}, who are rewarded for this by
the prestige of being member of an editorial board.
\end{itemize}
%Clearly, these features are related to each other: the absence of deadlines means that a paper can be revised as long as necessary to reach a desired level of quality. 

We would like to see these principles applied to good conference publications as well.  The easiest way to do this is to simply use the journal reviewing system for the conference.  That is: authors who submit to the journal, can indicate that they would like to have their article considered for presentation at the conference.  Submissions that are accepted, are then automatically accepted for the conference as well.  The authors can thus draw additional attention to their work.  Submissions may be considered not mature enough for publication in the journal, but interesting enough for presentation at the conference; these are again automatically accepted for the conference, without additional reviewing.

%We propose to apply these principles to conference submissions as well. Conceptually, hence, we would like to:
%\begin{itemize}
% \item allow submissions all year long;
% \item allow revisions all year long;
% \item accept papers all year long;
% \item create a permanent editorial board for ECML PKDD consisting of senior researchers.
%\end{itemize}
%Given that the papers would go through a reviewing process of journal quality, there is no reason not to  call the resulting set of papers ``journal articles''. Hence, submissions to ECML PKDD will ideally be published as journal articles.

However, to make this successful, we need to ensure that the revision and reviewing cycles are short. Indeed, the duration of journal reviewing is a major reason, for many researchers, for preferring conference publication.  We intend to achieve this by introducing some efficiency-increasing ideas from conference reviewing into the journal reviewing process.
%the popularity of conferences in computer science is after all that they allow researchers to publish their most recent results with relatively few delay.

These arguments could entail a solution in which ECMLPKDD  publishes all contributions in one or more scientific 
journals.  That is the VLDB model.  We believe this may be a too big step for ECMLPKDD at this point; it may require increasing the capacity of existing journals, or the creation of new ones.  Our initiative is intended to be a first, easily achievable  step in this direction.

%However, this may not be feasible for several reasons. First, existing scientific journals in machine learning and data mining do not typically accept the number of papers that is typically accepted at ECML PKDD. As an illustration, in 2011 ECML PKDD accepted approximately 120 papers; the machine learning journal in 2010 published only 60 articles in total. 

%Second, founding a new journal dedicated to ECML PKDD -- similar to the solution chosen by the VLDB conference -- requires resources that are not available to ECML PKDD organisers; convincing the community to allocate such resources  is expected to be hard as long as the value of a new system is not convincingly shown.

%Given these practical considerations, our proposal is discussed in the next section.

\section{Technical Details}

The proposed ECML PKDD submission system will have two tracks:
\begin{itemize}
 \item a proceedings track, which will have one deadline as usual, and whose proceedings will be published in lecture notes; a {\em program committee} will review these papers;
 \item a journal track, which will allow for submissions all year long, and whose articles will be published either in the machine learning or data mining journals, or in the lecture notes; a {\em guest editorial board} will review these papers.
\end{itemize}
At this moment, ECML PKDD's proceedings track will not be modified significantly. The main change is the introduction of a journal track. Only papers submitted to the journal track can enter the machine learning and data mining journals.  Papers that are not of journal quality, but of good conference quality, will be considered for inclusion in the proceedings even if submitted to the journal.  Accepted papers are immediately published online, either using the journal's {\em Online First} facility, or on the conference website (until the proceedings are there).\footnote{This idea is subject to approval from Springer.}  We are considering to mark papers that end up in the proceedings after journal quality reviewing.
The benefits for submitting to the journal track are the following:
\begin{itemize}
 \item papers can be submitted all year long; the possibility to resubmit gives authors a higher chance of presenting their work at ECML PKDD, while the repeated reviewing cycles should ensure the quality of the papers;
 \item multiple iterations of the same paper will be reviewed by the same reviewers, which should reduce the reviewing efforts; 
 \item papers will receive higher quality reviews, as the reviewing process is carried out by senior researchers;
 \item papers are available online almost immediately after acceptance (whether for journal or conference).
\end{itemize}
In detail, we propose the following setup.

\paragraph{Guest Editorial Board}
A Guest Editorial Board (GEB) will be appointed for the duration of one year, consisting of members of the editorial board of the machine learning and data mining journals, as well as additional senior researchers. GEB members will agree to the following:
\begin{itemize}
 \item timely reviewing of a limited number of articles that will be sent to them over the course of one year;
 \item monitoring of their performance.
\end{itemize}
Monitoring reviewer's performance is expected to be important to ensure a timely reviewing process. 
The process will however be lenient to reviewers as long as the chairs are notified in advance.  If reviewers do not meet their deadlines and do not notify the chairs well in advance, the assumption is that they will not be invited for next year's GEB.  In the following, asterisks indicate points that will count towards positive or negative evaluation of GEB members. 

Note that we explicitly do not limit the size of the GEB in advance. Even though we will start with a relatively small GEB, we intend to grow the editorial board as needed, depending on the popularity of the journal track. Our aim is to limit the number of articles that each GEB member has to review to at most 5.

\paragraph{Submission Procedure}
We envision the following submission procedure:
\begin{enumerate}
 \item There is a deadline on Sunday night, every two weeks.
 \item The list of submissions is sent immediately to the GEB on Monday morning. Each member will receive a personalized mail that contains the following:
\begin{itemize}
 \item an overview of the number of articles he/she reviewed;
 \item an average of the number of articles other editorial board members reviewed (once this number is high enough);
 \item a list of submitted articles (titles \& abstracts), each with a direct link on which the GEB member can click to show interest in this article.*
\end{itemize}
Furthermore, each GEB member will have the opportunity to post a comment on a submitted paper; these comments will not be shown to the authors, but can be seen as suggestions to the final reviewers; for instance, such a comment could be: ``this reminds me of the work by author A, the reviewers should check this''.*
 \item Based on these bids, the paper is manually assigned to reviewers on Friday, taking into account possible conflicts of interests. Papers that did not receive a sufficient number of bids, will be allocated a reviewer.
 \item The GEB member should accept or decline the task within the next week.*  %Refusals or lack of response will count negatively towards the member's performance, in particular when the member first shows interest and then refuses to review.
 \item The GEB member has 4 weeks for reviewing after being invited to review the paper, the deadline being on Friday.
 \item On Wednesday morning before the deadline, the GEB member is reminded by email of the approaching deadline. The GEB member can request an extension of one week by reacting to this mail before Friday.
 \item Any paper for which no response is received* is immediately allocated to another reviewer.  %The GEB member that did not meet his deadline will be evaluated negatively.
 \item The chairs aim to decide within one week whether the article {\em can} be published in the journal.  If positive, this decision is communicated to the editor-in-chief or a responsible action editor, who has to agree before the authors are notified. % To ensure the timeliness of the process and reduce the burden on the chairs, the input of 3 out of 4 chairs is sufficient to decide on the fate of a paper.
% the fate of the paper and communicate this to the authors.  If the chairs decide that the article can be published in the journal, however, they will forward this decision first to the editor-in-chief of the relevant journal, who will agree to this before the authors are notified. Hence, the decision to accept a paper in the journal may take longer than a rejection. 
Accepted papers will be published online as soon as possible.
\end{enumerate}
%Overall, in the best case this would provide a turn around time of only 6 weeks. 
Overall, this would result in a turn around time of only 6 to 8 weeks.  The VLDB experience suggests that this is realistic.

The last deadline that will allow for publication in the ECML PKDD 2013 journal track will be approximately 7 weeks before the deadline of the conference track. However, the chairs will not promise that reviewing will be finished within 7 weeks -- authors are recommended to submit sufficiently early.

Note that the ECML PKDD program chairs will decide on accepted papers both for the conference and the journal track. They will ensure that papers in both tracks have the desired quality; the editors-in-chief of the journals cannot autonomously decide to accept a paper in the conference track.

\paragraph{Ratings}
Reviewers will be given the opportunity to mark papers as follows:
\begin{enumerate}
 \item Accept immediately in the journal.
 \item Accept for the conference proceedings, but request revisions for the journal track.
 \item Request revisions and allow resubmission to the journal track.
 \item Reject.
\end{enumerate}
In case a paper is rated (2), the authors will have to indicate whether they intend to resubmit to the journal track. 
If not, the paper will enter the procedure for a final version in the conference proceedings.

In case a paper is rated (3), the authors are also allowed to resubmit their paper to the conference track instead of the journal track. After the last ECML PKDD 2013 journal track deadline has passed, this will be the only possible way to publish their results at ECML PKDD 2013. Such submissions will be reviewed by the GEB members that reviewed the original submission.

In case a paper is rated (4), the paper will also not be accepted to the conference track.

\paragraph{Format of Submissions}
Articles submitted to the journal track will have to be of journal quality. Hence, if experimental in nature, they will have to contain a sufficient number of experiments; if theoretical in nature, they will have to provide full proofs of their claims.

It will be possible to submit articles of any length to the journal track. However, only articles that adhere to a page limit of 20 pages in journal (DAMI/MLJ) style will be evaluated using the above described process, and will hence receive short review times. Articles beyond 20 pages will be treated as normal submissions to the corresponding journals; they cannot be forwarded to the proceedings and will typically not be reviewed in a short time.

%Appendices beyond 20 pages will be allowed, if they contain extended proofs or additional experimental validations. These appendices will not be published in lecture notes, but will be part of an accepted journal article. In case the appendices will not be published in lecture notes, the authors will be encouraged to publish the extended version in an open archive, such as arXiv. 
%There are two reasons for this limitation: first, a page limit on the core context of a journal article will speed-up the reviewing process; second, it allows to forward papers to lecture notes without increasing the length of the lecture notes too much. 

%\hendrik{Above, I would say max. 20 pages in journal style, excluding references; I would not say "given that...", because this practical reason is not a very good one in my opinion.  We want to keep reviewing fast, that is a second (and for me the first) reason!   Alternatively, if we restrict to 20 pages LNCS, which is shorter, we might consider making these articles "technical notes"?  I like also the arXiv alternative.}

By not imposing an overall page limit, we ensure that the quality of journal articles is not limited. At the same time, we do not believe it is reasonable to expect from reviewers that they provide high quality reviews for long papers within a short time.

%the articles will meet the standards of a journal article; at the same time, we believe that a good journal article should manage to convey its main message within 20 pages if full proofs or detailed experiments are excluded.

%At the same time, the number of pages that is submitted to the conference track will be limited to 12 pages; the main goal is also here to increase the quality of reviewing.

We will encourage all authors to publish their full versions also in arXiv/CoRR, and aim to have permissions for this from Springer.

\paragraph{Relationships to the Proceedings Track}
The proceedings track and the journal track clearly are not completely independent of each other. 
In particular, articles that are not revised in time for the journal track, may be resubmitted to the proceedings track. However, to reduce the reviewing load, such resubmissions from the journal track will be reviewed by the GEB members that reviewed the original submission. The number of reviews carried out by an GEB member will be recorded and taken into account when allocating proceedings submissions to reviewers; reviewers that reviewed a significant number of journal articles will not be asked to review proceedings submissions. Nevertheless, they will also be listed as members of the program committee; the program committee will automatically include all GEB members. 

\paragraph{Further Changes to the Proceedings Track}
There will be a separate deadline for the proceedings track.
The setup of the proceedings track, such as its acceptance rate and its reviewing procedures (whether or not to have author response, ...) are in principle independent from the setup of the journal track.
It would however be in the spirit of our proposal to accept a larger number of papers to the proceedings track and to reduce the selective pressure for this track. Other tracks, such as a demo track, an industrial track, or Nectar track, are under consideration.

%\section{Questions \& Answers}

%\begin{description}
%\item[\bf Q:] Will the perceived quality of the conference go down?  %(some expressed worries about the conference's citation index, whatever that may be)
%\item[\bf A:] The quality of the conference itself will go up.  The quality of the conference according to citation indexing may go down, as good papers will be in journals rather than conference proceedings. In the end, ECMLPKDD journal issues would need to be indexed in conference citation indexes, similar to how PVLDB issues need to be indexed. 
%However, more papers will be put in the proceedings.

%\item[\bf Q:] Will the quality of the conference proceedings go down? 
%\item[\bf A:] Yes, this is unavoidable. To turn the conference in a more inclusive meeting, more papers will be accepted, even if only as poster presentations. Nevertheless, we expect that good papers will continue to be submitted to the conference track; the conference track will even be more interesting, as it will provide the audience an opportunity to catch up with new trends earlier. 

%\end{description}

\section{Future Directions}
We envision that the system can evolve in several directions after ECMLPKDD'13.

\paragraph{A specialized journal}
ECMLPKDD can adopt the VLDB model and found a journal specific for ECMLPKDD, preferably open access in nature. This journal could be used for all full paper submissions; this journal may be indexed by conference indexing services to track the impact of ECMLPKDD. The present conference track reduces to a session for posters and short presentations on ongoing work.

\paragraph{Additional conferences join the system}
The ML and DMKD journals continue to receive submissions for ECMLPKDD, but other conferences start using ECMLPKDD's procedure, effectively establishing an ECMLPKDD reviewing system in collaboration with these journals.  At each resubmission the authors can indicate to which conference they wish to resubmit, allowing papers that not were not finished in time for ECMLPKDD to be considered in other conferences.

\paragraph{Additional journals join the system}
To increase the number of papers that ECMLPKDD can publish in established journals, it may be an option that more journals are allowed to join the system (a special purpose ECMLPKDD journal possibly being one of them). Based on the reviews, the chairs of the reviewing system allocate papers to journals; this is similar to conferences in bioinformatics, which nowadays often have special issues in many journals.

\paragraph{The Old System}
When our proposal is a failure, the journal track can be abolished and the conference track's acceptance rate can be decreased again.

\paragraph*{}
Key to our proposal is that it can evolve in each of these directions, based on the evolution of the field as a whole in the coming years.

%Summary:

%Pitfalls:

%Q: will the perceived quality of the conference go down?  (some expressed worries about the conference's citation index, whatever that may be)

%A: the quality of the conference itself will go up.  Quality of conference according to citation indexing may go down, as good papers will be in journals rather than conference proceedings.  However, more papers will be put in the proceedings (only 12p per paper; papers are presented as posters, so no artificial limit on how many can be accepted, but still included in full in proceedings).

\bibliographystyle{plainnat} 
\bibliography{refs}

\begin{thebibliography}{10}
\providecommand{\natexlab}[1]{#1}
\providecommand{\url}[1]{\texttt{#1}}
\expandafter\ifx\csname urlstyle\endcsname\relax
  \providecommand{\doi}[1]{doi: #1}\else
  \providecommand{\doi}{doi: \begingroup \urlstyle{rm}\Url}\fi

\bibitem[Birman and Schneider(2009)]{birman}
Ken Birman and Fred~B. Schneider.
\newblock Program committee overload in systems.
\newblock \emph{Communications of the ACM}, 52\penalty0 (5):\penalty0 34--37,
  2009.

\bibitem[Fortnow(2008)]{fortnow}
Lance Fortnow.
\newblock Viewpoint: Time for computer science to grow up.
\newblock \emph{Communications of the ACM}, 52\penalty0 (8):\penalty0 33--35,
  2008.

\bibitem[Grudin(2010)]{grudin}
Jonathan Grudin.
\newblock Technology, conferences, and community: Considering the impact and
  implications of changes in scholarly communication.
\newblock \emph{Communications of the ACM}, 54\penalty0 (2):\penalty0 41--43,
  2010.

\bibitem[Halpern and Parkes(2011)]{halpern}
Joseph~Y. Halpern and David~C. Parkes.
\newblock Journals for certification, conferences for rapid dissemination:
  Rethinking the role of journals in computer science.
\newblock \emph{Communications of the ACM}, 54:\penalty0 36--38, 2011.

\bibitem[Jagadish(2008)]{jagadish}
Hosagrahar~Visvesvaraya Jagadish.
\newblock The conference reviewing crisis and a proposed solution.
\newblock \emph{ACM SIGMOD Record}, 37:\penalty0 40--45, 2008.

\bibitem[Langford(2009)]{jl}
John Langford.
\newblock Summary of a discussion on future publication models \@ nips, 2009.
\newblock URL \url{http://hunch.net/?p=1086}.

\bibitem[LeCun(2010)]{lecun}
Yann LeCun.
\newblock A new publishing model in computer science, 2010.
\newblock URL \url{http://yann.lecun.com/ex/pamphlets/publishing-models.html}.

\bibitem[Naughton(2010)]{naughton}
Jeff Naughton.
\newblock Key note: {DBMS} research: First 50 years, next 50 years.
\newblock In \emph{IEEE International Conference on Data Engineering}, 2010.

\bibitem[Vardi(2009)]{vardi1}
Moshe~Y. Vardi.
\newblock Conferences vs journals in computing research.
\newblock \emph{Communications of the ACM}, 52\penalty0 (5):\penalty0 5, 2009.

\bibitem[Vardi(2010)]{vardi2}
Moshe~Y. Vardi.
\newblock Revisiting the publication culture in computing research.
\newblock \emph{Communications of the ACM}, 53\penalty0 (3):\penalty0 5, 2010.

\end{thebibliography}

\appendix
\newenvironment{citemize}{\begin{itemize}[noitemsep,topsep=0pt,parsep=0pt,partopsep=0pt]}{\end{itemize}}
\floatstyle{boxed}
\restylefloat{table} 

\section{Arguments against the Conference System}
\label{arguments}
The tables below summarize the arguments made against the conference system by several well-known computer scientists.

\begin{table}[h!]
\small
{\bf Reference}: \citet*{fortnow} \\
{\bf Credentials author}: 
\begin{citemize}
 \item Founding editor-in-chief of the ACM Transaction on Computation Theory
 \item Chair of ACM SIGACT
 \item Chair of the IEEE Conference on Computational Complexity 
\end{citemize}
{\bf Observations}:
\begin{citemize}
\item Rating researchers based on conferences is too random
\item The number of conferences is so large each individual one no longer brings together communities
\item Conferences can no longer accept all high-quality work
\item For papers on the margin, there is now a bias for ``safe'' papers (incremental and technical) and certain subareas (researchers from top CS departments dominate PCs and set the agendas)
\item Collaboration with researchers in other fields is hard due to having different publication procedures 
\item Papers get rejected due to simple misunderstandings
%- "By de-emphasizing their publication role, conferences can once again play their most important rule: 
 % Bringing the community together."
\end{citemize}
\hrule\vspace{0.3cm}
{\bf Reference}: \citet*{vardi1,vardi2} \\
{\bf Credentials author}: 
\begin{citemize}
 \item Editor-in-chief of the Communications of the ACM
\end{citemize}
{\bf Observations}:
\begin{citemize}
\item The computer science conference system compromises one of the cornerstones of scientific publication: peer review
\item The reviewing process performed by program committees is done under extreme time and workload pressure, and does not rise to the level of careful refereeing
\item The long turnaround times for journal articles is one important reason why computer science does not switch to a journal-oriented model; in other areas, the turnaround time is much lower
\item The conference-focused publication culture can not be separated from the sluggish journal 
  editorial process. Roles as editors and referees are not taken as seriously as PC memberships
\end{citemize}
\hrule\vspace{0.3cm}
{\bf Reference}: \citet*{grudin} \\
{\bf Credentials author}: 
\begin{citemize}
 \item Former editor-in-chief of the ACM Transactions on Computer-Human Interaction
 \item Associate Editor of ACM Computing Surveys
\end{citemize}
{\bf Observations}:
\begin{citemize}
\item A focus on conference publication has led to deadline-driven short-term research at the expense of journal publication, a reviewing burden that can drive off prominent researchers, and high rejection rates that favor cautious incremental results over innovative work
\item In other fields, journals focus on identifying and improving research quality; large conferences focus on community building: people don't retain quite the same warm feeling when their work is rejected
\item With pressure to reject $\approx75\%$ and differing views of what constitutes significant work, the minor flaws or literature omissions that inevitably accompany novel work become grounds for exclusion
\item It is increasingly difficult to evolve conference papers into journal articles. 
By Grudin's estimation, no more than 15\% of the work published in highly selective HCI conferences later appears in journals

\end{citemize}
\label{table:args1}
\caption{Arguments against the current conference-oriented publication model}
\end{table}

\begin{table}[h!]
\small

{\bf Reference}: \citet*{birman} \\
{\bf Credentials authors}: 
\begin{citemize}
 \item Former editor-in-chief of the ACM Transactions on Computer Systems
 \item Former editor-in-chief of Distributed Computing
 \item Associate editor-in-chief IEEE Security and Privacy
\end{citemize}
{\bf Observations}:
\begin{citemize}
\item Major conferences are overwhelmed by submissions
\item The conference system leads to more publications per researcher and per project, even though the aggregate scientific content of all these papers is likely the same as one journal article
\item Authors submit almost any paper to almost any conference, because acceptance will advance their research and career goals; rejection does them virtually no harm
\item The more innovative papers are the most likely to be either completely misunderstood or underappreciated by an increasingly error-prone process
\item There is a risk of a ``death spiral'' as senior people cease to review.  Young researchers often feel more comfortable identifying minor flaws and are less comfortable in declaring that work is more or less important
\end{citemize}

\hrule\vspace{0.3cm}
{\bf Reference}: \citet*{naughton} \\
{\bf Credentials author}: 
\begin{citemize}
 \item Keynote speaker at ICDE'10
 \item Associate Editor of VLDB Journal
 \item Associate Editor of ACM Transactions on Database Systems
\end{citemize}
{\bf Observations}:
\begin{citemize}
\item The combination of pressure to publish lots of papers, low acceptance rates, and bad reviewing, 
  is sucking the air out of our community
\item The conference system is often seen as an evaluation system for authors; in this sense, it can be compared to studying for and taking college entrance exams: `students' (authors) are not that interested in the questions, the `graders' (reviewers) are even less interested in the answers, no-one else is 
  interested either, caring only about the scores
\item Emphasis on paper count can be somewhat ameliorated by increasing acceptance rate: if it is easier 
  to publish papers, publishing lots of them will be perceived as less impressive; shifts the focus from 
  paper count to paper quality
\item SIGMOD 2010 received 350 submissions, 1 with uniform ``accept'' recommendations, 4 with average ``accept''
\item Receiving dysfunctional reviews begets writing dysfunctional reviews
\item Due to the absence of face-to-face PC meetings, there is less accountability pressure; there is fewer coaching and mentoring
\item If you get a `killer' reviewer, you are dead; so you resubmit until you have three non-killers
\end{citemize}
\vspace{0.3cm}
\caption{Arguments against the current conference-oriented publication model}
\label{table:args2}
\end{table}

\begin{table}[h!]
\small
{\bf Reference}: \citet*{jagadish} \\
{\bf Credentials author}: 
\begin{citemize}
 \item Founding editor-in-chief of the Proceedings of the Very Large Database Endowment (PVLDB)
 \item Program Chair of the International Conference on Intelligent Systems for Molecular Biology (ISMB), 2005
 \item Editor for the database section of the Computing Research Repository (CoRR)
\end{citemize}
{\bf Observations}:
\begin{citemize}
\item The number of submissions to many conferences has sky-rocketed, leading to
a downward spiral in reviewing quality and author satisfaction
\item The enormous size of program committees leads to huge variances in reviewing. An individual PC member
sees only a very small piece of the set of submissions; with
large PCs we have lost the normalization across accept decisions that a PC-based decision allows
\item Many rejected papers are resubmitted; the high {\em re}submission rate may be a leading cause of the high submission
rate
\item Two recent developments, ``rollover'' (resubmission ``with memory'' between conferences) and author feed-back, have received mixed appreciation
\item Jagadish proposes founding a ``Journal of Data Management Research'' that publishes papers for participating conferences 
\end{citemize}
\hrule\vspace{0.3cm}
{\bf Reference}: \citet*{lecun} \\
{\bf Credentials author}: 
\begin{citemize}
 \item General chair and organizer of the yearly ``Learning Workshop'' in Snowbird, Utah
 \item Associate editor of PLoS ONE
 \item Associate editor of the International Journal on Computer Vision 
 \item Program chair for NIPS '95, '94, '90, and many other conferences
\end{citemize}
{\bf Observations}:
\begin{citemize}
\item Conference reviews tend to favor papers containing incremental improvements on well-established methods, and tend to reject papers with truly innovative ideas
\item Reviewers get very little credit for doing a good job with reviews and can do a bad job with few adverse consequences
\item The current system is breaking down due to the enormous number of papers submitted and the impossibility of getting papers reviewed properly
\item The current evaluation system favors citations to well-known authors
\item Lengthy and faulty evaluations is what currently holds back the dissemination of good papers
\end{citemize}

 \caption{Arguments against the current conference-oriented publication model}
\label{table:args3}
\end{table}

\begin{table}[h!]
\small
{\bf Reference}: \citet*{jl} \\
{\bf Credentials author}:
\begin{citemize}
 \item Program chair of ICML'12
\end{citemize}
This text summarizes arguments of a discussion at NIPS'09, where {\bf observations} were:
\begin{citemize}
\item Reviewers are overloaded, boosting the noise in decision making
\item A new system should run with as little built-in delay and friction to the process of research as possible
\item It is bad to take double blind so seriously as to disallow publishing on arxiv or personal webpages
\item Any new system should appear to outsiders as if it is the old system, or a journal, because it is already hard enough to justify CS tenure cases to other disciplines
\item There should not be a big change with a complex bureaucracy, but rather smaller changes which are obviously useful or at least worth experimenting with
\end{citemize}

\hrule\vspace{0.3cm}
{\bf Reference}: \citet*{halpern} \\
{\bf Credentials authors}: 
\begin{citemize}
 \item Editor-in-chief of the Journal of the ACM
 \item Chairman of ACM Preprint
 \item Computing Research Repository (CoRR) coordinator
\end{citemize}
{\bf Observations}:
\begin{citemize}
\item Outside CS, there are two reasons for publishing in journals: certification and publicity
\item It is rare that conference reviewers review proofs as thoroughly as journal reviewers; for theoretical work certification in journals remains important
\item The approaches taken by SIGGRAPH and ICLP to directly publish conference papers in journals suffer from the same problems that conference pbulications suffer from: paper are subject to page restrictions and paper submission deadlines
\item The CS publication model complicates interdisciplinary research
\item Experimentation with publication models is needed
\item A system is proposed in which papers are submitted to public archives; journals are the main ``certification'' authorities that take papers from these archives; journals are much faster in reviewing papers. To achieve fast reviewing, page limits, better coordination between conferences and journals and imposing ``costs'' on certification are possible ideas
\end{citemize}

 \caption{Arguments against the current conference-oriented publication model}
\label{table:args4}
\end{table}

\begin{table}[h!]
\begin{small}
{\bf 1) Review quality is low and highly variable.}  %This is noticed in many areas of CS, not only in machine learning and data mining. 
The inclusion of inexperienced reviewers in program committees, as well as the need to review many papers within a short amount of time, leads to submissions being rejected for the wrong reasons.  Papers with good ideas get rejected because of minor flaws.  This very competitive context tends to favor incremental contributions.

{\bf 2) The number of papers submitted for reviewing is too large.} An important reason for this is clearly the need for researchers to publish to get funding (``publish or perish''). However, the situation is significantly worsened by the conference system: papers that are rejected in one conference will usually be resubmitted to another conference, where another set of reviewers is expected to evaluate the paper again; contributions are split over multiple (incremental) publications to meet page limits. 

{\bf 3) Acceptance for conferences is highly random.}  
Low acceptance percentages, combined with variable review quality, causes acceptance to be highly random.  This leads to frustration among authors, and decreases the quality of the conference, not only because less good work may be presented, but even more because it misses out on good work.  

{\bf 4) Journal papers are often unnecessarily long.}
Their extension over a (series of) conference paper(s) often consists of additions made only to make the paper ``sufficiently different'' from the previous version(s).  Often, all crucial information was already in the conference version(s) (otherwise it would not have been accepted there).

{\bf 5) Conference papers in CS are undervalued outside and inside the field.}  Our system is not only misunderstood by many people outside the field (including, but not limited to, people who evaluate our research), but also increasingly within the field.  Some areas in CS, such as bio-informatics, statistical learning, \ldots are moving towards the standard model, due to the influence of biologists, physicists, \ldots  This biases selection of papers at conferences, as well as selection of candidates for postdoc and tenure track positions.

{\bf 6) The status of CS journals is too low.} Journals in computer science do not publish the work with the highest impact: by some estimates only 15\% of results in computer science ends up in journals. The slow reviewing cycles lead to low impact factors, as few articles that build on earlier journal work will be published within the 2 years that are needed to increase the most well-known impact factors. Consequently, journals do not have the same visibility as conferences.

{\bf 7) Authors are faced by a ``conference versus journal'' dilemma.} Authors have the choice between: 
submitting directly to a journal (less visibility), submitting only to a conference, not to a journal (disadvantageous in countries where evaluation is based on journal publications), submitting first to a conference, then (an extended version) to a journal (problematic because journal articles must ``significantly extend'' conference papers)

{\bf 8) Deadline-driven research leads to worse papers}, which are not as polished as one might hope.  One might argue that these will be rejected anyway, but due to the variability in reviewing this is not necessarily so.  In any case, submission of unpolished papers causes a waste of time and effort on the authors', reviewers', and possibly readers' side; furthermore, such papers will typically lead to incremental follow-up papers, which increase the reviewing load later on as well.

{\bf 9) Proliferation of conferences.}  When there is a higher demand for papers to be published than there are conferences, the number of separate conferences grows. People lack the time and/or money to visit all the conferences.  As a result, the role of conferences as places where the whole community gets together is diluted.

{\bf 10) Researchers are increasingly overloaded with reviewing tasks.} This discourages the best researchers in the field to stay actively involved. As a result, reviewing is often delegated to younger researchers, even PhD students.  This lowers the quality of the reviews to an unacceptable level. %(Ref: Naughton)
\end{small}

\caption{A summary of arguments against the current conference-oriented publication model}
\label{table:args}
\end{table}

\end{document}